# Comparative study of electron and hole doped High-Tc compounds in pseudogap regime: LDA+DMFT+$\Sigma_\mathbf{k}$ approach.

I.A. Nekrasov[a*1], E.E. Kokorina[a], E.Z. Kuchinskii[a], Z.V. Pchelkina[b], M.V. Sadovskii[a]

[a]*Institute for Electrophysics, Russian Academy of Sciences, Ural Branch, Amundsena str. 106, Ekaterinburg, 620016, Russia*

[b]*Institute for Metal Physics, Russian Academy of Sciences, Ural Branch, S. Kovalevskoj 18, Ekaterinburg 620041, Russia*

**Abstract**

Pseudogap regime for the prototype high-$T_c$ compounds hole doped $Bi_2Sr_2CaCu_2O_{8-x}$ (Bi2212) and electron doped $Nd_{2-x}Ce_xCuO_4$ (NCCO) is described by means of novel generalized LDA+DMFT+$\Sigma_\mathbf{k}$ approach. Here conventional dynamical mean-field theory (DMFT) equations are supplied with additional (momentum dependent) self-energy $\Sigma_\mathbf{k}$. In the present case $\Sigma_\mathbf{k}$ describes non-local dynamical correlations induced by short-ranged collective Heisenberg-like antiferromagnetic spin fluctuations. Material specific model parameters of two neighboring CuO2 layers of Bi2212 and single $CuO_2$ layer of NCCO were obtained within local density approximation (LDA) and constrained LDA method. We show that Fermi surface in presence of the pseudogap fluctuations have perfectly visible "hot-spots" for NCCO while in Bi2212 there is just rather broad region with strong antiferromagnetic scattering. Results obtained are in good agreement with recent ARPES and optical experiments.

Keywords: Pseudogap, LDA+DMFT, Quasiparticle bands, Fermi Surface;

## 1. Introduction

Pseudogap phenomena are observed for normal underdoped phase of different high-Tc cuprates. Among others the hole doped compound $Bi_2Sr_2CaCu_2O_{8-x}$ (Bi2212) is one of the most studied experimentally [1]. On the other hand electron doped high-Tc prototype system is $Nd_{2-x}Ce_xCuO_4$ (NCCO) [1]. In accordance with common understanding Mott insulators under moderate doping become strongly correlated metals. Thus at finite doping (up to optimal doping) high-$T_C$ cuprates are typical strongly correlated metals. Also quasi two-dimensional nature of these compounds is well known. The Hubbard model supposed to be a relevant model for strongly correlated metals. To take pseudogap and correlation effects into account simultaneously we solve the Hubbard model with calculated material specific parameters for CuO2 layer within LDA+DMFT+$\Sigma_\mathbf{k}$ approach.[2]

## 2. Computational method

In this work electronic structure of Bi2212 and NCCO was investigated within recently proposed generalized LDA+DMFT+$\Sigma_\mathbf{k}$ computational scheme [2]. This scheme has advantage to combine first principle density functional theory in local density approximation (LDA) [3] with dynamical mean-field theory (DMFT) [4] to solve correlation problem for real materials. To overcome local nature of DMFT (for example, for quasi two dimensional problem) we supply it with external momentum dependent self-energy $\Sigma_\mathbf{k}$ [5]. Using our general approximation, namely ignoring interference effects between DMFT Hubbard interaction and interactions responsible for $\Sigma_\mathbf{k}$, we can keep conventional DMFT or LDA+DMFT [6] set of equations for any type of physics $\Sigma_\mathbf{k}$ reflects [5].

As a first stage we perform LDA band structure calculations. Both compounds have ideal tetragonal bcc crystal lattice with space symmetry group I4/mmm (for Bi2212 see Refs. [7], for NCCO Ref. [8]). Main structural motif for Bi2212 compound is two $CuO_2$ layers displaced close to each other in the unit cell. Using the crystal structure data we have done LDA calculations of electronic band structure within the linearized muffin-tin orbital (LMTO) basis set [9]. Obtained band structures are in agreement with previous results of Refs. [7,10] and Ref. [11] for Bi and Nd compounds correspondingly. To calculate hopping integral values for Bi system Wannier functions projecting method [12] in the LMTO framework [13] was applied. Hopping integrals in Nd compounds were obtained using so called NMTO method [14]. Results of both methods were compared with each other and agree well for the same compounds [15]. The values of hopping integrals between $x^2-y^2$ orbital of different Cu-sites are listed in the Table 1 for both compounds. The values of local Coulomb interaction $U$ for $x^2-y^2$ orbital were obtained in constrained LDA method [16] (Table 1).

[1]*Corresponding author. Tel.: +7-343-267-8823; fax: +7-343-267-8794; e-mail: nekrasov@iep.uran.ru.

To study the "antiferromagnetic" scenario of pseudogap formation in cuprates [17] **k**-dependent self-energy $\Sigma_\mathbf{k}$ describing non-local correlations induced by (quasi) static short-ranged collective Heisenberg-like antiferromagnetic (AFM) spin fluctuations is included [18]. These fluctuations are predominantly determined by scattering with reciprocal vector $\mathbf{Q}=(\pi,\pi)$ and are characterized by energy scale $\Delta$ (pseudogap potential) and correlation length $\xi$.

Pseudogap potentials $\Delta$ were calculated as described in Ref. [5] and are listed in Table 1. The values of correlation length $\xi$ were taken to be equal to 5 lattice constants for Bi2212 [19] and 50 lattice constants for NCCO [20] as typical experimental values. Hole doping level $\delta$ is 15% in Bi system and electron doping in Nd system is 10%. To solve the effective single impurity problem in LDA+DMFT+$\Sigma_\mathbf{k}$ equations the numerical renormalization group (NRG) [21] is applied.

**3. Results and discussion**

On the left side of Fig. 1 LDA+DMFT+$\Sigma_\mathbf{k}$ Fermi surface (FS) for Bi2212 is presented. Close to the borders of BZ one can see significant FS "destruction" because of pseudogap fluctuations. Also shadow FS is observed for our LDA+DMFT+$\Sigma_\mathbf{k}$ results. Right side of Fig. 1 displays FS for NCCO. Comparing left and right panels of Fig. 1 one can conclude that FS "destruction" in NCCO happens not close to BZ border but in the so called "hot-spots". The same FS shapes are observed experimentally for both Bi [22] and Nd [23] compounds. Our results agree well with presented experimental data (see Fig. 1 lower line). Such a difference can be explained from material specific point of view. Namely FS of NCCO has more curvature and thus at the BZ boundary remains nearly noninteracting one. While Bi2212 FS comes to BZ border much closer to the $(\pi,0)$ point. Because of that in Bi2212 "hot-spots" are not seen in Bi2212. They are smeared out by strong pseudogap scattering processes near $(\pi,0)$ point.

Fig. 2 displays LDA+DMFT+$\Sigma_\mathbf{k}$ ARPES spectra along 1/8 of noninteracting FS from antinodal (lower curve) to nodal point (upper curve). Left panels correspond to ARPES spectra of Bi2212 obtained theoretically (upper line) and experimental data [24] (lower line). Right panels show the same quantity for NCCO. In general for both compounds in antinodal point quasiparticles are well-defined - sharp peak close to the Fermi level. Towards nodal point we obtained damping of the quasiparticle peak and its shift to higher binding energies. Similar behavior was observed experimentally [23,24]. However there are some differences between these compounds. As we said before "hot-spots" for NCCO are closer to the BZ center. This fact gives another origin of the peaks seen. Namely, for Bi2212 nodal quasiparticles are formed by low energy edge of pseudogap while for NCCO they are formed by higher energy pseudogap edge. Also in NCCO there is no bilayer splitting effects seen for Bi2212 (left part of Fig. 2).

In Fig. 3 we show spatial dependence of quasiparticle static scattering rate that is just the value of self-energy imaginary part taken at the Fermi level. For this quantity we found the same tendency as before: "hot-spots" are more pronounced for NCCO than for Bi2212 and are closer to BZ center. Nevertheless experimental maximal scattering values for both compounds are approximately the same [23,24]. As to theoretical results one can conclude that for Bi system calculated value of the pseudogap potential is slightly smaller than in nature but for Nd compound it is quite overestimated. However as one can see in Figs. 1 and 2 this is not very crucial for FS and ARPES shapes. But one should mention here that in Ref. [24] authors tried to map their data on to some model self-energy while in Ref. [23] it is just half-width on a half height. This fact can cause the discrepancy.

On Fig. 4 real part of optical conductivities for NCCO (left panel) and Bi2212 (right panel) are presented in comparison with experimental data. To calculate theoretical curve our recent generalization of DMFT+$\Sigma_\mathbf{k}$ with respect to two particle properties was applied [25]. Here we can say that qualitatively our theoretical curve for NCCO with calculated $\Delta=0.36$ eV (Fig. 4, solid line) agrees reasonably with experiment [26]. But again we find calculated pseudogap value to be about 2 times overestimated. That was already mentioned in previous paragraph. To improve the agreement we also calculated optical conductivity for experimental value of $\Delta=0.2$ eV [26] (Fig. 4, dashed line). The possible source of these discrepancies could also arise from underestimation of the value of on-site Coulomb interaction U that is calculated in our work. Concerning Bi2212 optical conductivity (Fig. 1, right panel) one can point out that there is no particular structure neither in theory nor in the experimental data [27]. Again for Bi2212 agreement between experimental and theoretical curves is reasonable.

Journal of Physics and Chemistry of Solids     3

## 4. Summary

To summarize our comparative study the difference in the physical quantities discussed (FS, ARPES, static scattering rate) can be explained just by the differences in nonintersecting electronic band structures. Strong correlation effects included here via novel generalized LDA+DMFT+$\Sigma_\mathbf{k}$ approach are rather similar for both Nd and Bi compounds, though obviously it is important for correct physics. Especially remarkable are evident "hot-spots" in NCCO FS. Concerning pseudogap features one can conclude that pseudogap effects are a significantly stronger in NCCO system. It follows for example from model parameters calculated and also from optical conductivity. In NCCO pseudogap is very well developed and in Bi2212 experimental optical conductivity is pretty featureless.

Table 1. Calculated energetic model parameters (eV).

|        | t      | t'    | t"     | t'''    | tBS   | U    | Δ    |
|--------|--------|-------|--------|---------|-------|------|------|
| Bi2212 | -0.627 | 0.133 | 0.061  | -0.015  | 0.083 | 1.51 | 0.21 |
| NCCO   | -0.44  | 0.153 | -0.063 | -0.0096 | ---   | 1.1  | 0.36 |

### Acknowledgments

This work is supported by RFBR grants 05-02-16301, 05-02-17244, RAS programs "Quantum macrophysics" and "Strongly correlated electrons in semiconductors, metals, superconductors and magnetic materials", Dynasty Foundation, Grant of President of Russia MK.2242.2007.2, interdisciplinary UB-SB RAS project.

### References

1] A. Damascelli, Z. Hussain, Z.-X. Shen, Rev. Mod. Phys. **75**, 473 (2003); J.C. Campuzano, M.R. Norman, M. Randeria, In "Physics of Superconductors", Vol. II, ed. K. H. Bennemann and J. B. Ketterson (Springer, Berlin, 2004), p. 167-273; J. Fink, *et al.*, cond-mat/0512307; X.J. Zhou, *et al.*, cond-mat/0604284.
2] Kuchinskii E.Z. *et al.*, JETP **131**, 908 (2007);
3] W. Kohn and L. J. Sham, Phys. Rev. **140**, A1133 (1965); L. J. Sham and W. Kohn, Phys. Rev. **145**, 561 (1966).
4] A. Georges, *et al.*, Rev. Mod. Phys. **68**, 13 (1996).
5] E. Z. Kuchinskii, I. A. Nekrasov, M.V. Sadovskii, JETP Lett. **82**, 198 (2005); M.V. Sadovskii, *et al.*, Phys. Rev. B **72**, 155105 (2005). E.Z.Kuchinskii, I.A.Nekrasov, M.V.Sadovskii Low Temp. Phys. **32**, No.4/5, 528 (2006).
6] K. Held, *et al.*, Psi-k Newsletter **56**, 65 (2003).
7] M. Hybertsen and L. Mattheiss, Phys. Rev. Lett. **60**, 1661 (1988).
8] Kamiuama T. Physica C **229**, 377 (1994); see also Ref. [11].
9] O. K. Andersen, Phys. Rev. B 12, 3060 (1975); O.K. Andersen and O.Jepsen, Phys. Rev. Lett 53, 2571 (1984).
10] J. M. Tarascon, *et al.*, Phys. Rev. B **37**, 9382 (1988); S. A. Sunshine, *et al.*, Phys. Rev. B **38**, 893 (1988).
11] Massida S. *et al.*, Physica C **157**, 571 (1989); Agraval B.K. *et al.*, Phys. Rev. B **43**, 1116 (1991); Matsuno S., Kanimura H., J. of Superconductivity **7**, 517 (1994).
12] N. Marzari and D. Vanderbilt, Phys. Rev. B 56, 12847 (1997); W. Ku, *et al.*, Phys. Rev. Lett. **89**, 167204 (2002).
13] V. I. Anisimov, *et al.*, Phys. Rev. B **71**, 125119 (2005).
14] O.K. Andersen and T. Saha-Dasgupta, Phys. Rev. B **62**, R16219 (2000); O.K. Andersen, *et al.*, Psi-k Newsletter **45**, 86 (2001); O.K. Andersen, T. Saha-Dasgupta, and S. Ezhov, Bull. Mater. Sci. **26**, 19 (2003).
15] Korshunov M.M., *et al.*, JETP (3): 559 (2004).
16] O. Gunnarsson, *et al.*, Phys. Rev. B 39, 1708 (1989).
17] M.V. Sadovskii, Physics-Uspekhi **44**, 515 (2001); cond-mat/0408489.
18] J. Schmalian, D. Pines, B.Stojkovic, Phys. Rev. B **60**, 667 (1999).E. Z. Kuchinskii, M. V. Sadovskii, JETP **88**, 347 (1999).
19] For review seeRef. 17.
20] Gukasov A.G., *et al.*, Solid State Comm. **95**, 533 (1995); Motouama E.M. *et al.*, Nature **445**, 186 (2007).
21] R. Bulla, A.C. Hewson, and Th. Pruschke, J. Phys. Cond. Mat. **10**, 8365 (1998); R. Bulla, Phys. Rev. Lett. **83**, 136 (1999).
22] Borisenko S.V., *et al.*, Phys. Rev. Lett. **84**, 4453 (2000).
23] Armitage N.P., *et al.*, Phys. Rev. Lett. **71**, 014517 (2005).
24] A. Kaminski, *et al.*, Phys. Rev. B **88**, 257001 (2002).
25] E.Z. Kuchinskii, I.A. Nekrasov, M.V. Sadovskii, Phys. Rev. B 75 115102 (2007).
26] Onose Y., *et al.*, Phys. Rev. Lett. **87**, 217001 (2001).
27] Quijada M.A. , *et al.*, Phys. Rev. B **60**, 14917 (1999).

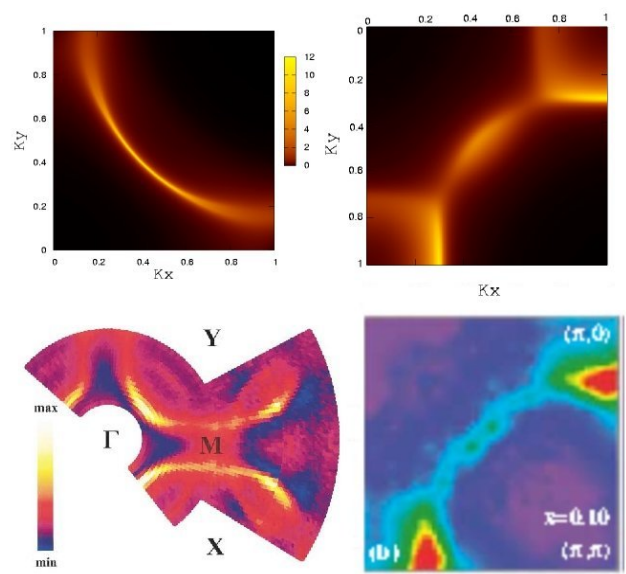

Fig. 1. LDA+DMFT+$\Sigma_\mathbf{k}$ Fermi surface (¼ of BZ) for Bi2212 (left panels) and NCCO (right panels). Theoretical results (upper line) are contour plot of Green function imaginary part -1/πImG($\mathbf{k}$,ω=0). Lower line shows experimental data for Bi2212 [22] and NCCO [23].

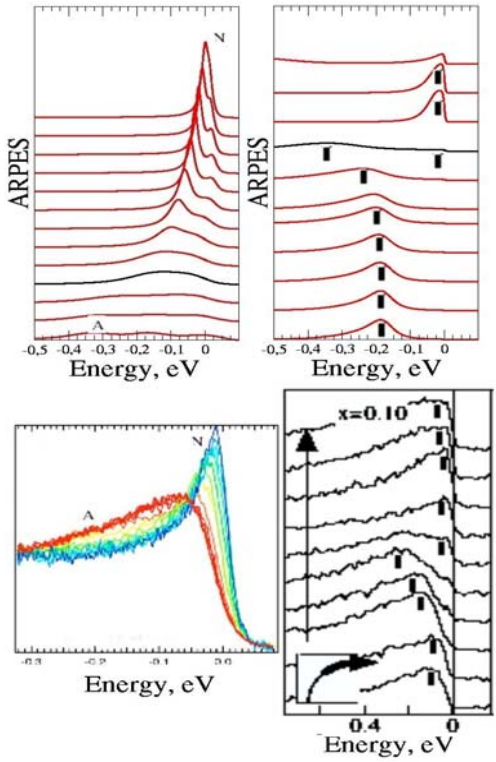

Fig. 2. LDA+DMFT+$\Sigma_k$ calculated ARPES spectra for Bi2212 (upper left panel) and NCCO (upper right panel) along of noninteracting FS in 1/8 of BZ. Corresponding theoretical full Green function imaginary parts $-1/\pi \text{Im} G(\mathbf{k},\omega)$ are multiplied with Fermi function at T=255K (the temperature of NRG calculations). Lower line shows experimental data for Bi2212 [24] and NCCO [23].

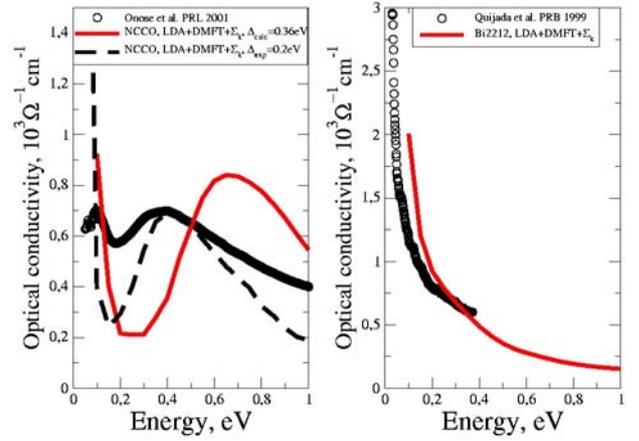

Fig. 4. Comparison of LDA+DMFT+$\Sigma_k$ calculated optical conductivity spectra for NCCO (left panel) with experimental data [26] (circles). Solid line – theoretical results for calculated pseudogap value 0.36 eV (dashed line corresponds to experimental pseudogap value 0.2 eV) On the right panel there is the same quantity but for Bi2212 and experiment of Ref. [27].

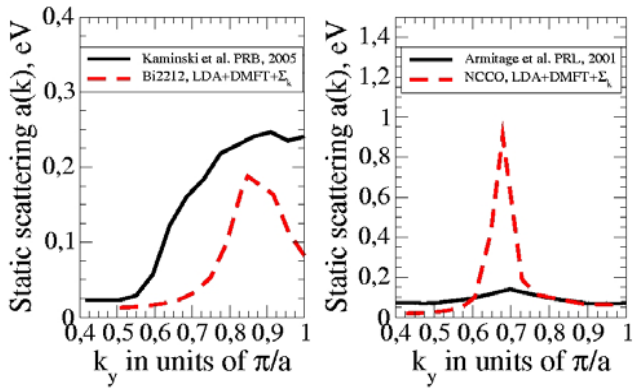

Fig. 3. Comparison of LDA+DMFT+$\Sigma_k$ calculated (dashed) and experimental (black) spatial dependencies of static scattering a(k) for Bi2212 (left) [24] and NCCO (right) [23] along of noninteracting FS in 1/8 of BZ.